# Related Rhythms: Recommendation System To Discover Music You May Like

**Rahul Singh**[*]
Language Technology Institute
Carnegie Mellon University
Pittsburgh, PA 15213
`rahuls2@cs.cmu.edu`

**Pranav Kanuparthi**[*]
Electrical and Computer Engineering
Carnegie Mellon University
Pittsburgh, PA 15213
`pkanupar@andrew.cmu.edu`

## Abstract

Machine Learning models are being utilized extensively to drive recommender systems, which is a widely explored topic today. This is especially true of the music industry, where we are witnessing a surge in growth. Besides a large chunk of active users, these systems are fueled by massive amounts of data. These large-scale systems yield applications that aim to provide a better user experience and to keep customers actively engaged. In this paper, a distributed Machine Learning (ML) pipeline is delineated, which is capable of taking a subset of songs as input and producing a new subset of songs identified as being similar to the inputted subset. The publicly accessible Million Songs Dataset (MSD) enables researchers to develop and explore reasonably efficient systems for audio track analysis and recommendations, without having to access a commercialized music platform. The objective of the proposed application is to leverage an ML system trained to optimally recommend songs that a user might like.

## 1  Introduction

The magnitude of data available in the MSD [1] not only opens up possibilities for ML, but also warrants applications which leverage distributed workflows with efficacious data-processing and analysis. Song-recommendation systems are one such application capable of suggesting tracks in real-time. The work presented in this paper aims to address the following question -

*"Given a set of songs that we know a user likes, can we predict a disjoint set of songs that is likely to match the user's song choice?"*

We briefly discuss the methodology, techniques and metrics used in building the ML pipeline. This is followed by an explanation of the software infrastructure deployed and the computational challenges faced. The final section expounds on the results, analysis and inferences drawn.

### 1.1  Task

As mentioned in the previous section, we aim to recommend songs to a user given input song(s). We do this by selecting a subset of audio tracks from the MSD that are classified as being similar to the input. From this subset, we compute the count of similar artists and determine the top-n frequently occurring similar artists, where 'n' is a tuned hyperparameter. Finally, we present the songs retrieved by these chosen artists belonging to this subset of the original dataset as the recommended songs.

---

[*] Equal contribution

## 1.2 Dataset

In the ML pipeline, we relied on the Million Song Dataset, which comprises audio features and metadata extracted from a million music tracks. A complete description of the features along with an example of the metadata schema can be found at the MSD web-page *linked here*. The MSD was constructed using several other datasets as a part of a collaborative effort between Echo Nest and LabROSA, created under an NSF grant using a cluster of community contributed datasets such as SecondHandSongs dataset and musiXmatch dataset. A few noteworthy statistics of the dataset are highlighted in Table 1.

Table 1: MSD Statistics

| Attribute | Magnitude |
|---|---|
| Size | 275 GB |
| Unique Artists | 44,745 |
| Dated Tracks | 515,576 |
| Songs | 1,000,000 |

The dataset can be downloaded as h5 files with the metadata and audio features. As seen in Figure 1, the audio features are mel-frequency cepstral coefficients (MFCC) which are essentially spectrograms of the recordings with each time-step having twelve frequencies. There are a total of 54 features. Among other attractive attributes, the data includes features such as 'danceability', 'loudness', 'key confidence', 'bars confidence', 'artist familiarity' and 'artist location'.

## 2 Methodology

In this section, we describe the overall ML pipeline used from end-to-end and also explain the various stages involved in developing the proposed application.

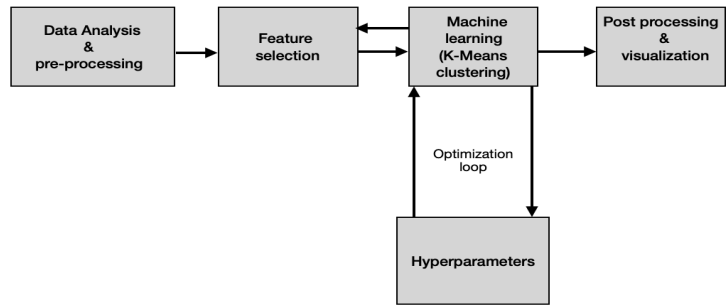

Figure 1: High-level block diagram of the ML Pipeline

### 2.1 Data Retrieval and Analysis

We used Amazon Web Services (AWS) as the cloud infrastructure to host the system. Data acquisition and pre-processing involved retrieving all 280GB of data and converting the files from HDF5 to a suitable format for analysis, which took about three days. The dataset residing on an AMI (Amazon Machine Image) was first obtained on an S3 bucket, after which we set up an Amazon EC2 instance from where we processed the data and converted it into csv files. These files were uploaded back to S3, and finally we were able to interact with the data using an Elastic Map-Reduce (EMR) instance running PySpark. Once the data was loaded, we performed some analysis of the features and computed some statistics associated with the features. Analyzing the data helped in data engineering and feature selection.

## 2.2 Feature Selection

After assessing the various features of the dataset, we decided to condense the dataset by considering only the mean of some features rather than the entire sequence that was originally present and by dropping features with zero variance and those with sparse entries. Next, based on our experimental results, we decided to drop most features which solely consisted of strings from the condensed dataset and considered features only with numerical values for the machine learning algorithm. While these features were used to train the clustering model, we relied on textual features in the subsequent phase such as artist_terms - a list of strings describing the genre of the song - to generate recommendations and unique strings (track_ID and artist_ID) to identify individual audio tracks within clusters.

## 2.3 Machine Learning

We used K-means to cluster our data such that it is capable of extracting the necessary information from a given subset of music tracks. Figure 2 depicts the process adopted for clustering.

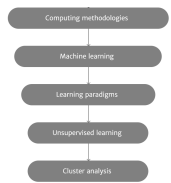

Figure 2: Breakdown of the unsupervised learning process

We used a clustering algorithm, an unsupervised learning mechanism for our task as we hypothesized that this would enable the algorithm to classify songs based on a diverse set of similarities rather than a single attribute like genre.

## 2.4 Model Optimization

We performed feature selection, training and tuned hyperparameters in a cyclic manner in order to optimize our model. For each value of k, we used silhouette scores to measure how similar an object is to the cluster it was assigned compared to the other clusters. We also verify that the recommended songs are similar to the input song and not something absurd. After each phase of experiments, the results were analyzed and the performance was improved in a subsequent iteration by pruning ineffective features and appending new features to our dataset. To choose the right number of clusters for our K-means model, we first conducted grid search on 10% of the MSD, and using the results of this experiment we selected an initial batch of features to use. We grew the dataset to 25% and repeated the process, before finally conducting random search on the entire data in order to choose the optimal number of clusters.

## 3 Computation

The system was hosted on Amazon EMR which made it seamless to deploy clusters by providing a secure, reliable and flexible cloud based big data platform and built-in tools. Data pre-processing was done using Python while the rest of the code-base used PySpark. The main libraries used were pyspark.ml, pyspark.sql, pandas, numpy and s3fs. Initially, we planned to launch a 3-node EMR cluster with an m4.xlarge lead node and two r4.2xlarge core nodes and had estimated an overall expense of $50 with 5 hour daily usage for a period of one month. Most of the computational challenges were faced in the early stages of the pipeline. It took over two days to retrieve the entire dataset in the pre-processed form. First, we loaded the entire dataset and discovered that we could not interact with the HDF5 file format directly using PySpark. So we converted the files to npy format only to realize that there were major formatting issues and a large chunk of data was lost in the conversion. We stopped this data-load midway and converted the data to csv files instead, and made sure to parallelize the process and drop some unnecessary features at this stage - which made the loading much faster. At this point, we also had to update our cloud infrastructure by switching entirely to m5.xlarge nodes, which helped us stay within the $100 budget. Apart from the data retrieval stage, the run-time estimates made were accurate, with the major time consumption occurring in the training and hyperparameter-search stages of the ML model. Using the entire dataset, model training took about an hour to complete on average. The inference and visualization code ran in approximately one and five minutes respectively. Running grid-search was not an option as it would take days to complete, so we relied on random search which we ran on a limited number of configurations. In retrospect, capacity planning was a great learning experience. Given the size of the dataset and the formatting of the data, performing analysis and making the right estimates - of storage and compute requirements as well as expenses - proved crucial.

## 4 Results and Analysis

This section briefly describes the experimental setup as well as some of the results obtained, analysis conducted and inferences made. Following the iterative development strategy characteristic of our ML-pipeline, we were able to make incremental updates based on inferences made after running experiments, and this helped boost our results.

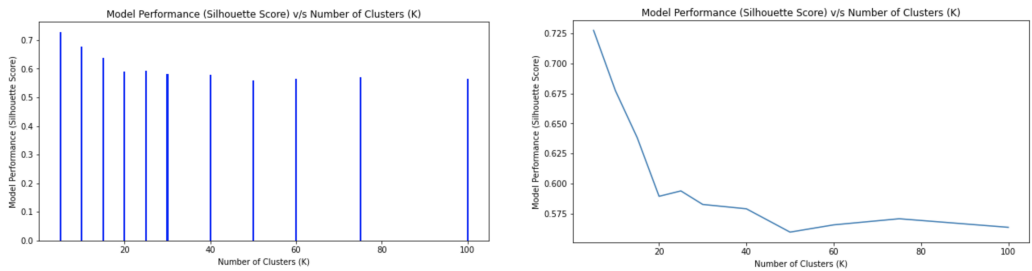

Figure 3: Visualization of model performance based on silhouette score v/s number of clusters

We conducted experiments by varying the value of 'k' - the number of clusters and obtained the highest silhouette score for k = 5. We found that the silhouette scores decreased almost monotonically as we increased the values of k, as seen in the figure.

During feature selection, we identified some features which were negatively impacting the model performance and introducing a skew in the predicted cluster centroids. Some of the pruned features included attributes from songs metadata, segments/sections/loudness/bars confidence. We also dropped the song_length feature, which was negatively influencing our results and we concluded that it was not useful in terms of the task, which is also in line with the logical interpretation - since song-similarity is unlikely to be influenced by the length of the tracks.

Although the silhouette score was best when k = 5, we found that the song recommendations were not satisfactory. After analysis, we found that the best trade off between song recommendations and silhouette score was at k = 20.

| Input (song- artist) | Genre | Recommended songs (song-artist) | Genre |
|---|---|---|---|
| **Melodia Do Mal - OIO AI** | soft rock<br>pop rock<br>indie rock<br>post rock<br>rock | Rip-Off Poppoff (2004 Digital Remaster) -TC Matic | new wave<br>pop rock<br>new beat<br>rock<br>alternative rock |
| | | Please Come (This Mystery Album Version) - Nichole Nordeman | soft rock<br>pop rock<br>ccm<br>j pop<br>british pop |

Figure 4: A truncated example of song recommendations for k = 20

For better understanding, a few simplified instances of system output have been included in the Appendix at the end, where you can see the input song and the corresponding recommendations produced. These include a result from the final model as well as an initial system result which produced sub-par recommendations.

## Acknowledgments

We would like to thank Virginia Smith and Ameet Talwalkar for their constant guidance, for providing us useful suggestions and also for giving us an opportunity to work on this task. We are grateful to Baljit Singh and Ignacio Maronna for their constructive feedback and support.

# Appendix

## A  Simplified end-to-end system results

| Input song | Genre | Recommended songs | Genre |
|---|---|---|---|
| **Friday Night Blues** | country rock<br>country<br>folk<br>world<br>rock | Early Grave | oi<br>hardcore punk<br>alternative rock<br>country rock<br>rock |
| | | What's Up With Trapper Records | oi<br>hardcore punk<br>alternative rock<br>country rock<br>rock |
| | | Siete Vidas | boogaloo<br>salsa<br>doo-wop<br>rock n roll<br>latin jazz |
| | | Think I Wanna Die (1) | indie rock<br>british pop<br>pop rock<br>indie pop<br>rock |
| | | #21 (Album Version) | electronica<br>techno<br>electronic<br>trance<br>intelligent dance music |
| **Rain When I Die** | grunge<br>alternative metal<br>heavy metal<br>hard rock<br>alternative rock | Women III | new wave<br>europop<br>new romantic<br>electronic<br>pop rock |
| | | Long Road | trip hop<br>future jazz<br>downtempo<br>breakbeat<br>acid jazz |
| | | I Put A Spell On You | country rock<br>rock<br>rockabilly<br>blues<br>funk |
| | | The Great Drive By | trip hop<br>future jazz<br>downtempo<br>breakbeat<br>acid jazz |
| | | Travelin' Band | roots rock<br>heartland rock<br>rock<br>country<br>americana |
| | | W (2000 Digital Remaster) | art rock<br>experimental<br>symphonic rock<br>experimental rock<br>free jazz |

Figure 5: Two instances of song recommendations produced by the system. These examples are simplified and scaled down to a single song input to represent the basic functionality of our system. The first instance is from the latest version of the system while the second instance is drawn from an initial end-to-end system validation experiment run during the nascent stages of the overall pipeline.